\documentclass[aps,pra,reprint,superscriptaddress]{revtex4-2}
\usepackage{xcolor}
\usepackage{graphicx}
\usepackage{bbm}
\usepackage{braket}
\usepackage{amsmath}
\usepackage{svg}
\usepackage{siunitx}
\usepackage{relsize}
\usepackage{verbatim}

\usepackage{fancyhdr}
\begin{document}

\title{\textbf{Dynamic distributed quantum sensing of radio-frequency fields via time-bin entanglement } 
}%
\author{Vedansh Nehra}
\thanks{These authors contributed equally to this work. Corresponding author: erin.sheridan.1@us.af.mil}
\affiliation{Technergetics LLC, Utica, New York, 13502, USA}

\author{Richard Birrittella}
\affiliation{Booz Allen Hamilton, Rome, New York, 13441, USA}

\author{Benjamin Malia}
\affiliation{Technergetics LLC, Utica, New York, 13502, USA}

\author{Nicholas J. Barton}
\affiliation{Murray Associates of Utica, Utica, New York, 13502, USA}

\author{Christopher C. Tison}
\affiliation{Air Force Research Laboratory, Rome, New York, 13441, USA}

\author{James Schneeloch}
\affiliation{Air Force Research Laboratory, Rome, New York, 13441, USA}

\author{David Hucul}
\affiliation{Air Force Research Laboratory, Rome, New York, 13441, USA}

\author{Benjamin Kyle}
\affiliation{Air Force Research Laboratory, Rome, New York, 13441, USA}

\author{Erin Sheridan}
\thanks{These authors contributed equally to this work. Corresponding author: erin.sheridan.1@us.af.mil}
\affiliation{Air Force Research Laboratory, Rome, New York, 13441, USA}


\begin{abstract}
We propose a novel framework for discrete-variable (DV) distributed quantum sensing, wherein traditional polarization-based probes are replaced with time-bin entangled qubits. We show that the intrinsic temporal structure of time-bin Bell states can be utilized as a tunable, built-in two-time differential sampling reference for sensing radio-frequency (RF) fields. Unlike existing protocols based on polarization Bell states, which are limited to static measurements, our proposed architecture enables the coherent mapping of dynamic RF phase signals onto static quantum optical phases. By matching the time-bin separation to the RF or intermediate frequency (IF) half-period (antipodal sampling), we theoretically demonstrate that our protocol provides a 6 dB sensitivity enhancement over one-sample static polarization encodings while preserving the 3 dB per-pair entanglement advantage over the separable standard quantum limit benchmark. The performance enhancement is validated with Monte Carlo simulations. This approach provides a solution for RF sensing in DV photonic quantum networks, offering a clear path toward quantum-enhanced and frequency-agile distributed sensor arrays and broader utilization of time-domain quantum probe encodings.
\end{abstract}

\maketitle
\thispagestyle{fancy}


\section{\label{sec:intro}Introduction}
Distributed quantum sensing (DQS) is an often-cited application of quantum networks \cite{Zhang2021} wherein quantum resources, including entanglement and squeezing, are used to enhance measurements beyond the standard quantum limit (SQL)\cite{giovannetti2004, Giovannetti2006}. As quantum networking technologies mature, DQS is seeing increased attention in both the experimental \cite{Malia2022, Xia2023, Zhang2026} and theoretical domains \cite{Czekaj2015, zhuang2019, BirrittellaAlsingGerry2021, VanMilligen2024, Bringewatt2024, Guo2026}. Photonic DQS proposals and demonstrations have leveraged both continuous variable (CV) \cite{Zhuang2018, grace2020, Xia2020, xia2021, Zhang2021, hao2022} and discrete variable (DV) quantum probe states \cite{zhao2021, Kim2024, Kim2025}.

For radio frequency (RF) sensing, specifically, CV DQS networks utilizing squeezed light and multipartite entanglement have demonstrated phase estimation enhancements beyond the SQL \cite{Xia2020, grace2021, Sun2022, Li2023, Li2024}. CV DQS networks rely on balanced homodyne detection, which inherently requires the co-propagation of a local oscillator (LO) with the squeezed entangled probe. Selecting the measurement quadrature (e.g. phase) demands stable probe-LO phase locking. This is achievable in real-world networks \cite{Chapman2023, chapman2025, Verclas2026}, but requires careful engineering and sophisticated phase stabilization infrastructure. The sensitivity of squeezing to loss \cite{Frascella2021} also motivates the exploration of alternative DQS solutions that are more robust against degradation for real-world operation.

DV-DQS relies on single photon detection in place of balanced homodyne detection. Though single photon detection often increases the cost, size, weight and power of the network infrastructure, it carries the benefit of not requiring a co-propagating and phase-locked LO for quadrature locking. Recently, Kim \emph{et al.}\cite{Kim2024} demonstrated $2.2\mathrm{\;dB}$ sensitivity enhancement relative to the standard quantum limit (SQL) when measuring four phases with two photons, with polarization Bell states. In their work, they move beyond the established mode-entangled and particle-entangled (MePe) DV-DQS probe states \cite{Liu2021}
\begin{align}
    \ket{\Psi_{\mathrm{MePe}}} = \frac{1}{\sqrt{2}} \big( \bigotimes^d_{j=1}\ket{H_j}^{\otimes N/d} + \bigotimes^d_{j=1}\ket{V_j}^{\otimes N/d} \big)
\end{align}
 
\noindent where $H(V)$ denotes a horizontal (vertical) polarization, $N$ denotes the total number of probe photons, and $d$ is the number of sensing nodes. $N/d$ is the number of photons at each sensor node. For MePe states, $N = m \times d$ must hold for an integer $m\geq 1$. Achieving the highly sought Heisenberg scaling with MePe states is resource-intensive, in that $m$-photon GHZ states are required. Such states are difficult to generate for large $d$ and are also prohibitively sensitive to loss.

The probe state in Ref.~\cite{Kim2024} is instead a coherent superposition of $N$-photon two-mode polarization entangled states
\begin{multline}\label{eq:pol_bell}
\ket{\Psi_d^N}=\frac{1}{\sqrt{2d}}\sum_{j=1}^d \big( \ket{H_j}^{\otimes N/2}\ket{H_{j+1}}^{\otimes N/2} + \\\ket{V_j}^{\otimes N/2} \ket{V_{j+1}}^{\otimes N/2} \big)
\end{multline}

\noindent where $N/2$ photons are distributed to two adjacent nodes. The photon number at each sensing node is $N/2$, which introduces the constraint that $N$ is an even number. A two-photon probe state $\ket{\Psi_4^2}$ with $N=2$ and $d=4$ is distributed through a beamsplitter network (BSN) and measures four unknown phases $\phi_j$, $j=1,2,3,4$. 

\begin{multline}
\ket{\Phi_{a,b}}=\frac{1}{\sqrt{2}} \big( \ket{H_aH_b} + \ket{V_aV_b} \big) \\
    \xrightarrow[]{BSN}\ket{\Psi_4^2} = \frac{1}{2} \big( \ket{\Phi_{1,2}} + \ket{\Phi_{2,3}} + \ket{\Phi_{3,4}} + \ket{\Phi_{4,1}} \big)
\end{multline}

They estimated the average $\hat{\phi}$ of four spatially distributed phases:
\begin{equation}
    \hat{\phi}=\sum_{j=1}^4 \frac{\phi_j}{4}.
\end{equation}
Classical probes limit the estimation variance of this average phase to the standard quantum limit scaling $\mathrm{Var}(\hat{\phi})_{\mathrm{SQL}} \propto 1/(d N)$, whereas the Bell state probe achieves Heisenberg scaling with respect to the photon number, yielding a variance scaling of $\mathrm{Var}(\hat{\phi})_{\mathrm{Bell}} \propto 1/(d N^2)$. An MePe state utilizes global $d$-body entanglement to achieve a scaling of $1/d^2$ with respect to the number of sensor nodes, subject to the rigid constraint that $N=d$. The Bell state probe relies on distributed bipartite entanglement, which limits its spatial node scaling to $1/d$, but decouples the photon resource from the node count ($N \geq d$) and reduces the network's sensitivity to photon loss.

In that architecture, the estimated phase $\hat{\phi}$ corresponds to a static polarization rotation $\theta$ imparted on the probe by a set of waveplates. It is a configurable angle of an optical element placed in the optical beam path and not a property of any external field. Related work has used multimode $N00N$ states \cite{Hong2021, Kim2025}, but the measured observable remains a static polarization rotation.

While external fields can couple to a polarization qubit by inducing polarization rotations, the class of observables accessible to static-$\theta$ protocols is limited to signals whose temporal variation is slow compared to the entanglement distribution and coincidence accumulation timescales. Coherent, frequency-resolved or sub-microsecond observables fall outside of this class because the polarization qubit encodes information at a single time per photon, with no internal temporal structure available for sub-cycle demodulation. Time-varying parameters can be tracked only by repeated static measurements with the measured parameter held quasi-stationary between events, which restricts this protocol to DC or low-frequency phenomena. Extending such architectures to dynamic, high-frequency RF fields remains a significant challenge.

To apply this distributed framework to the sensing of time varying observables, such as RF electromagnetic fields, one must overcome the static-observable limitation by appending external time-domain machinery to the polarization Bell state for demodulation, such as RF-locked pulse pumping, polarization switching, or post-selection by arrival-time tagging. This bottleneck motivates the exploration of alternative DV probe states capable of coherently recovering the carrier phase of an RF signal. In this work, we show that these hardware limitations can be overcome by extending the DV-DQS framework from the static-parameter regime to the time-varying regime, leveraging time-bin encoded probe states \cite{marcikic2004}. We show that time-bin encoded Bell states achieve the same $3\;\mathrm{dB}$ entanglement advantage in precision over separable probe states, while enabling an additional $6\;\mathrm{dB}$ precision advantage when the time-bin separation $\Delta \tau$ is properly set and locked. We evaluate the impact of the main limitations of our method, including inefficient microwave-optical transduction, showing how they can be overcome in real-world network operation.

\section{\label{sec:proposal}Proposed method}

\subsection{Probe state preparation}
The RF carrier phase $\psi$ is a temporal property of the electromagnetic field. To extract it coherently from a photon, one requires a probe state analogous to Eq.~\ref{eq:pol_bell} but with temporal structure:
\begin{equation}
\begin{aligned}
\ket{\Psi_d^N}=\frac{1}{\sqrt{2d}}\sum_{j=1}^d \big( \ket{E_j}^{\otimes N/2}\ket{E_{j+1}}^{\otimes N/2} +  \\ \ket{L_j}^{\otimes N/2} \ket{L_{j+1}}^{\otimes N/2} \big)
\end{aligned}
\end{equation} 
where $\ket{E}$ and $\ket{L}$ denote early and late temporal modes separated by time delay $\Delta \tau$. These time-bin Bell states can be generated using an unbalanced Mach-Zehnder interferometer (UMZI) with a differential propagation time of $\Delta \tau$, which in principle can be tuned.

\begin{figure*}[htbp]
  \centering
  \includegraphics[width=\linewidth]{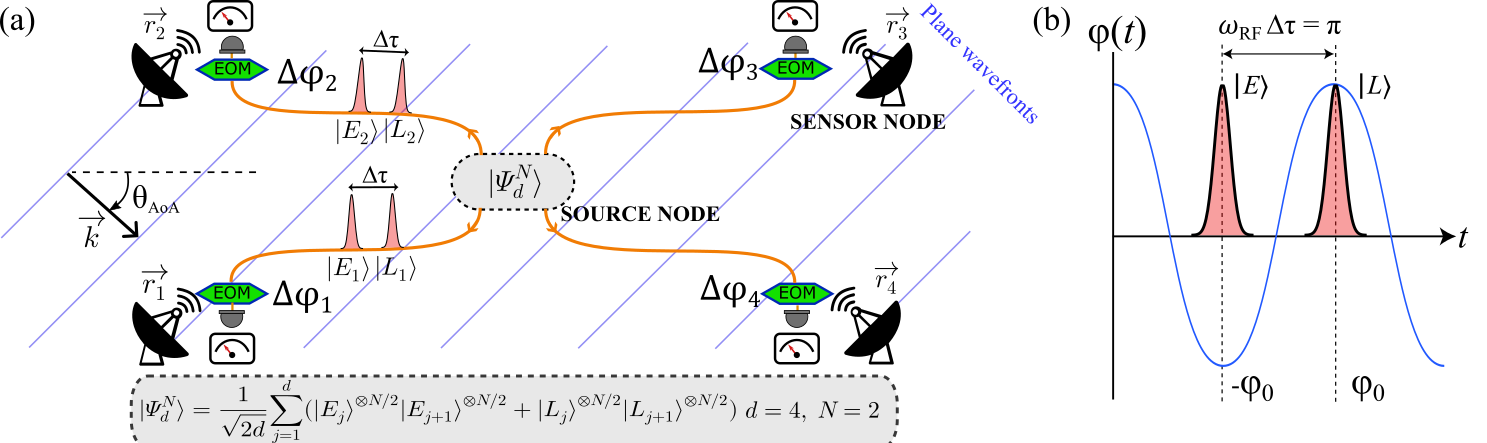} 
  \caption{(a) Simplified network diagram. An incident RF field, with angle of arrival $\theta_{\mathrm{AoA}}$ of its wave-vector $\vec{k}$ with respect to the sensor baseline, is sensed by four sensor nodes at positions $\vec{r}_j$. A time-bin Bell state (here for $N=2\;, d=4$) is distributed from a central source node to the four remote sensor nodes. Electro-optic modulator (EOM)-based RF-to-optical transduction imprints the local RF field phase $\psi_j$ onto the probe state before measurement, inducing an optical phase shift $\Delta \varphi_j$. (b) Coherent sampling of the RF waveform using an Early-Late time-bin separation $\Delta \tau$. }
  \label{fig:intro_diagrams}
\end{figure*}

In this modified protocol, as before, the Bell state is distributed to four sensor nodes via a BSN:
\begin{equation}
\begin{aligned}\label{eq:time_Bell}
\ket{\Phi_{a,b}}=\frac{1}{\sqrt{2}} \big( \ket{E_aE_b} + \ket{L_aL_b} \big) \\
    \xrightarrow[]{BSN}\ket{\Psi_4^2} = \frac{1}{2} \big( \ket{\Phi_{1,2}} + \ket{\Phi_{2,3}} + \ket{\Phi_{3,4}} + \ket{\Phi_{4,1}} \big)
\end{aligned}
\end{equation}
The UMZI can have its static phase $\phi_{\mathrm{UMZI}} \in [0, \pi]$ adjusted to select $\ket{\Phi^-}$ instead of $\ket{\Phi^+}$. Likewise, $\ket{\Psi^{+/-}}$ Bell states can be generated by applying a local bit-flip operation ($\hat{\sigma}_x$) to one of the spatial modes prior to network distribution. Physically, this time-bin permutation can be realized using an active electro-optic switch and a calibrated delay loop to deterministically swap the early and late modes. Alternatively, as we detail below, one can first generate a polarization Bell state and then convert to the time-bin basis \cite{nehra2026}, granting the network configurable access to the full suite of Bell states.

\subsection{Radio frequency transduction}
Each sensor node $j \in [1,..., d]$  includes an antenna with gain $G$ and length $h_e$, coupled to an electro-optic modulator (EOM) with a half-wave voltage $V_{\pi}$. An incoming plane wave at carrier frequency $\omega_{\mathrm{RF}} $ and amplitude $E_0$ produces the voltage
\begin{align}
V_j(t)=G h_e E_{\mathrm{RF}}(\mathbf{r}_j, t)=G h_e E_0\cos(\omega_{\mathrm{RF}}t+\psi_j)
\end{align}
where $\psi_j= -\mathbf{k}\cdot \mathbf{r}_j$ is the carrier phase of the RF signal with wavevector $\mathbf{k}$ at position $\mathbf{r}_j$. The instantaneous phase imparted on a photon as it passes through the EOM at time $t$ is then
\begin{align}
    \varphi_j(t) = \frac{\pi V_j(t)}{V_{\pi}} = \varphi_0\cos(\omega_{\mathrm{RF}}t+\psi_j),
\end{align}
assuming a modulation depth $\varphi_0 = \pi G h_e E_0/V_{\pi}\ll 1$ for small signals and linear EOM operation.

The early mode $\ket{E}$ of the time-bin qubit probe arrives at the EOM at time $t_e$, and the late mode $\ket{L}$ arrives at time $t_e + \Delta \tau$. Each mode picks up an instantaneous optical phase at its own arrival time. After both modes pass through the EOM, the relative phase acquired by the qubit is 
\begin{align}
    \Delta \varphi_{j} \equiv \varphi_j (t_e + \Delta \tau) - \varphi_j(t_e). 
\end{align}
If we define $\Delta \tau$ such that 
\begin{align}\label{eq:demod_identity}
\omega_{\mathrm{RF}}\Delta \tau = \pi,
\end{align}
the late mode will sample the RF signal exactly one-half period after the early bin samples it, so that
\begin{equation}\label{eq:early_late_gain}
\begin{aligned}
    \Delta \varphi_j = \varphi_0 [\cos(\omega_{\mathrm{RF}}t_e+ \pi + \psi_j) - \cos(\omega_{\mathrm{RF}}t_e+\psi_j)]\\
    = -2\varphi_0\cos(\omega_{\mathrm{RF}}t_e + \psi_j).
\end{aligned}
\end{equation}
We can see that the RF carrier phase has now been mapped onto a constant optical phase - the relative phase between the $\ket{E}$ and $\ket{L}$ bins within the time-bin qubit. The time-bin separation $\Delta \tau$ acts as a local oscillator of the RF carrier frequency demodulation. The factor of $2$ that arises is coherent doubling due to antipodal sampling, which yields a four-fold Quantum Fisher Information (QFI) increase, or equivalently a factor-of-two reduction in the Cram\'er--Rao standard deviation $6.0\;\mathrm{dB}$. 

Treating $\Delta \tau$ as a controllable parameter at the central source renders the time-bin qubit a frequency-agile optical probe that can span a broad range of RF frequencies. Inserting a tunable delay element into the source and detection UMZIs, or alternatively incorporating a tunable RF local oscillator at each node, expands the network's functionality with minimal additional overhead. In practice, tuning optical delays may limit tunable range of operation. We propose an RF mixing scheme that enables $\Delta \tau$ tuning without changing the optical delay line. See Section~\ref{sec:network} for more details.

\begin{figure*}[htbp]
  \centering
  \includegraphics[width=\linewidth]{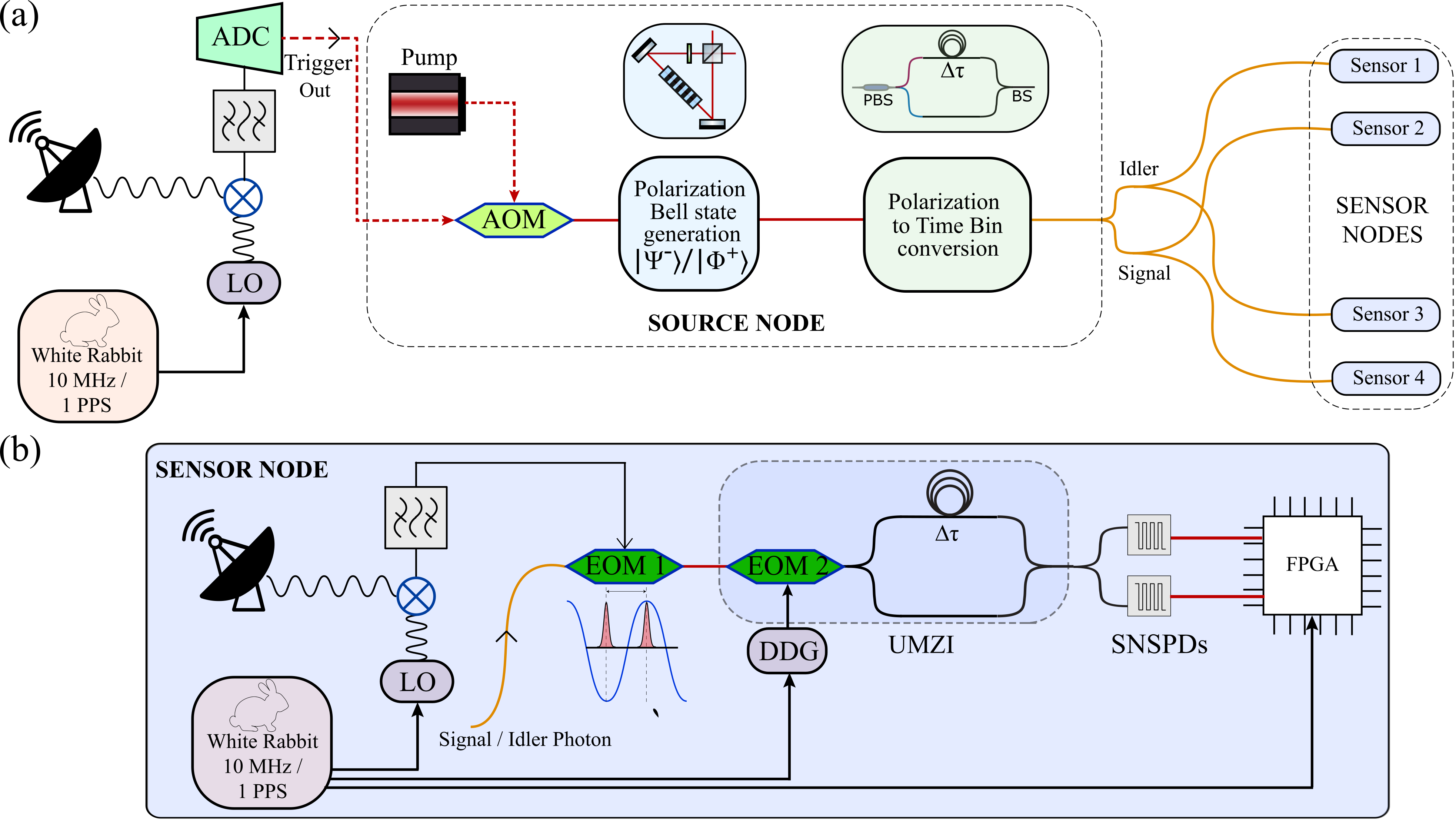} 
  \caption{Proposed network architecture operating at baseband. (a) The source node generates a polarization Bell state via a standard Sagnac interferometer, then converts it to the time-bin basis. An antenna is included to establish an absolute phase reference for the RF signal of interest. In this scenario, the RF signal is mixed with a local oscillator (LO). The baseband intermediate frequency signal is routed to an analog  to digital converter (ADC), which is used to trigger an acousto-optic modulator (AOM) to meet the quadrature locking condition, and a phase-locked loop is used to stabilize the RF phase as needed.  Output signal and idler time-bin photons each impinge upon a beamsplitter, so that the Idler travels to Sensors 1 or 3, and the Signal travels to Sensors 2 or 4. A White Rabbit (WR) fan-out at the source conditions the LO phase and distributes $10\;\mathrm{MHz}$ and $1\;\mathrm{PPS}$ signals throughout the network. (b) Each sensor node has its own antenna to receive the RF signal, and mixes the signal to baseband using an LO that is conditioned on the distributed WR clock signal. The baseband signal is filtered and sent to an electro-optic modulator (EOM 1) for transduction. EOM 1 imprints the RF phase onto the time-bin photon. The photon carrying the RF phase information is routed through an unbalanced Mach-Zehnder Interferometer (UMZI) then detected with a superconducting nanowire single photon detector (SNSPD). EOM 2 is set by a digital delay generator (DDG) to apply the fast phase shift $\xi_j$ that sets the analyzer measurement basis on the time-bin qubit. Detection events are recorded on a field programmable gate array (FPGA) and synthesized across the network for measurement.}
  \label{fig:network}
\end{figure*}

\subsection{Unitary phase evolution of the probe}

The action of the EOM on the probe photon at sensor node $j$ is diagonal in the time-bin basis, since the Early and Late modes acquire the instantaneous optical phases $\varphi_E^j = \varphi_j(t_e)$ and $\varphi_L^j = \varphi_j(t_e + \Delta \tau)$, respectively. The single-node evolution operator is hence
\begin{align}
    \hat{U}_j = e^{i\varphi_E^j} \ket{E}\bra{E} + e^{i \varphi_L^j}\ket{L} \bra{L}
\end{align}
which we decompose into a global phase and a $\hat{\sigma}_z$ rotation:
\begin{align}
    \hat{U}_j = e^{i\bar{\varphi}_j}\exp{\big(- i \frac{\Delta \varphi_j}{2} \hat{\sigma}_z\big )}
\end{align}
with $\bar{\varphi}_j=(\varphi_E^j+\varphi_L^j)/2$ , $\Delta \varphi_j = (\varphi_L^j-\varphi_E^j)$ and $\hat{\sigma}_z= \ket{E}\bra{E} - \ket{L}\bra{L}$. Imposing the demodulation identity (Eq.~\ref{eq:demod_identity}) yields 
\begin{align}
    \varphi_L^j = \varphi_0\cos(\omega_{\mathrm{RF}}t_e+\psi_j+\pi)=-\varphi_E^j
\end{align}
so that the global phase becomes zero. The EOM action then reduces to a pure time-bin $\hat{\sigma}_z$ rotation
\begin{align}
      \hat{U}_j = \exp{\big(- i \frac{\Delta \varphi_j}{2} \hat{\sigma}_z\big )}, \;\;\;\;\Delta \varphi_j = -2\varphi_0 \cos(\omega_{\mathrm{RF}}t_e+\psi_j).
\end{align}
The cancellation of the common-mode phase $\bar{\varphi}_j$ is a direct consequence of the antipodal sampling dictated by Eq.~\ref{eq:demod_identity}. The Early and Late temporal modes acquire equal and opposite phases, leaving only the qubit-frame rotation that carries the RF signal phase $\psi_j$. The unitary operator for two-node phase evolution is a tensor product 
\begin{align}
    \hat{U}_{ij}=\hat{U}_i \otimes \hat{U}_j= \exp[-i (\frac{\Delta \varphi_i}{2} \hat{\sigma}_z^i+\frac{\Delta \varphi_j}{2} \hat{\sigma}_z^j)].
\end{align}
For a Bell state (Eq.~\ref{eq:time_Bell}) distributed between nodes $i$ and $j$, the evolved state after EOM-based sensing is 
\begin{equation}
\begin{aligned}
    \hat{U}_{ij} \ket{\Phi^+}_{ij} = \frac{1}{\sqrt{2}}(e^{-i(\Delta \varphi_i+ \Delta\varphi_j)/2}\ket{EE}+ \\ e^{i(\Delta \varphi_i+\Delta \varphi_j)/2}\ket{LL} \big).
\end{aligned}
\end{equation}
If we define the relative phase $\Theta_{ij}^{\Phi}$:
\begin{align} \label{eq:rel_phase_phi}
\Theta_{ij}^{\Phi}\equiv \Delta \varphi_i+\Delta\varphi_j,
\end{align}
we can write the resulting state as 
\begin{align}
    \hat{U}_{ij} \ket{\Phi^+}_{ij} = \frac{1}{\sqrt{2}}\big(\ket{EE}+e^{i \Theta^{\Phi}_{ij}}\ket{LL} \big).
\end{align}
If we instead probe with $\ket{\Psi^-}$, the relative phase becomes $\Theta_{ij}^{\Psi}$
\begin{align} \label{eq:rel_phase_psi}
\Theta_{ij}^{\Psi}=\Delta \varphi_i-\Delta\varphi_j
\end{align}
and
\begin{align}
    \hat{U}_{ij} \ket{\Psi^-}_{ij} = \frac{1}{\sqrt{2}}\big(\ket{EL}-e^{i \Theta^{\Psi}_{ij}}\ket{LE} \big).
\end{align}
We can select the desired Bell state at the probe source depending on which pairwise observable we want to measure: $\ket{\Phi^+}$ corresponds to measurement of the sum of the phases, whereas $\ket{\Psi^-}$ corresponds to measurement of the phase difference. Access to both observables enables the implementation of a variety of phase estimation methods for different RF sensing use cases. Substituting Eq.~\ref{eq:early_late_gain} into Eqs.~\ref{eq:rel_phase_phi}, \ref{eq:rel_phase_psi} yields
\begin{align}
    \Theta_{ij}^{\Phi}=- 4 \varphi_0\cos \big( \omega_{\mathrm{RF}}t_e  + \frac{\psi_i + \psi_j}{2} \big) \cos \big( \frac{\psi_i - \psi_j}{2}\big) \\
      \Theta_{ij}^{\Psi}=  4 \varphi_0\sin \big( \omega_{\mathrm{RF}}t_e  + \frac{\psi_i + \psi_j}{2} \big) \sin \big( \frac{\psi_i - \psi_j}{2}\big).
\end{align}
The phase difference term $(\psi_i - \psi_j)$ appears in the second factor and is independent of $t_e$, whereas the phase sum term $(\psi_i + \psi_j)$ sits inside of the first term with $t_e$. The implication of this asymmetry is that the $t_e$ dependence of the phase difference in $\ket{\Psi^-}$ is limited to a common-mode envelope $\sin \big( \omega_{\mathrm{RF}}t_e  + \frac{\psi_i + \psi_j}{2} \big)$ that depends on the sum of the phases at nodes $i,j$. In contrast, the phase sum term in $\ket{\Phi^+}$ sits within this $t_e$-dependent envelope, and an absolute phase reference would be required to extract its value.   The asymmetry between sum and difference observables in their phase reference requirements has, to our knowledge, not yet been noted in prior DV-DQS literature.

While not strictly required for phase-difference extraction, quadrature locking of the $\ket{E}$ mode arrival time makes sum phase extraction possible, and also allows for optimization of the operating point with respect to the sinusoidal envelope. Specifically, setting and locking $t_e$ such that $\omega_{\mathrm{RF}}t_e = \pi/2$  yields, in the small angle regime $|\psi_j|\ll1$ :
\begin{align}
\Theta_{ij}^{\Phi} \approx  2\varphi_0(\psi_i + \psi_j) \\
\Theta_{ij}^{\Psi} \approx2\varphi_0(\psi_i - \psi_j).
\end{align}
Both observables enjoy the same maximum sensitivity coefficient $2\varphi_0$ at this locking point. It is important to note that arrival time locking is not a given. Locking $t_e$ to an RF carrier phase requires a shared RF phase reference, and may not be possible in real-world scenarios where the RF signal emitter is not cooperative. This requirement can be circumvented through post-processing, at the cost of adding classical demodulation backend infrastructure that we are striving to avoid.

\subsection{Photon detection and measurement}
After the probe state $\ket{\Psi}$ interacts with the phase encoding $\hat{U}(\boldsymbol{\phi})$ and evolves to $\hat{U}(\boldsymbol{\phi})\ket{\Psi}$, it is detected by a set of projectors ${\hat{\Pi_l}}$. The measurement at each node is performed by a local UMZI that is matched in arm-delay to the source UMZI ($\Delta \tau$), with a controllable phase $\xi_j$ on the long arm, followed by single-photon detection at the two output ports. The UMZI maps the two-dimensional time-bin qubit onto three resolvable output time slots: an early slot (early bin via short arm), a center slot (early bin via long arm, coincident with late bin via short arm), and a late slot (late bin via long arm). Only the central slot exhibits interference, since only there do two indistinguishable amplitudes overlap.

Detection in the central slot implements a projective measurement of the time-bin qubit in the equatorial basis set by $\xi_j$. With each 50:50 UMZI beam splitter contributing an amplitude factor $1/\sqrt{2}$, the Kraus operator for a central-slot detection at output port $a \in \{+, -\}$ is
\begin{equation}
\hat{K}_{a,j} = \tfrac{1}{2}\ket{c}\!\left(\bra{L} + a\,e^{i\xi_j}\bra{E}\right),
\label{eq:kraus}
\end{equation}
where $\ket{c}$ denotes the central-slot output mode. The associated
positive operator-valued measure (POVM) element $\hat{E}_{a,j} =
\hat{K}_{a,j}^\dagger \hat{K}_{a,j} = \tfrac{1}{2}\,
\ket{\chi_{a,j}}\!\bra{\chi_{a,j}}$ projects, with central-slot post-selection probability $\tfrac{1}{2}$, onto the equatorial states
\begin{equation}
\ket{\chi_{a,j}} = \tfrac{1}{\sqrt{2}}\!\left(\ket{L} + a\,e^{-i\xi_j}\ket{E}\right).
\end{equation}
A pair of detectors at the two UMZI output ports thus realizes a complete equatorial-basis measurement on the central-slot subspace. Scanning the UMZI phase $\xi_j$ rotates the measurement basis around the equator of the time-bin Bloch sphere. For the bipartite coincidence measurement, both photons must be detected in their respective central slots, which occurs with joint post-selection probability $\tfrac{1}{2} \times \tfrac{1}{2} = \tfrac{1}{4}$. Note that while we model a passive 50:50 UMZI here, an active electro-optic switch driven by a digital delay generator could be used to deterministically route early and late time-bins into their respective arms, eliminating the central-slot post-selection penalty at the expense of additional switch insertion loss. In Fig.~\ref{fig:network}(b), EOM 2 or an additional EOM could be used for active photon routing in addition to UMZI phase control. All Fisher-information bounds in this work are conditioned on the number of registered central-slot coincidence events $N_c$. Under this resource convention, passive interferometric splitting losses change the event rate but not the conditional information per registered coincidence. A complete source-to-estimate comparison must additionally include transmission, detector, switching, and post-selection efficiencies

The coincidence amplitude for ports $(a_i, a_j)$ on the evolved state $\hat{U}\ket{\Psi^-}$ is
\begin{equation}
\langle c,c|\,\hat{K}_{a_i,i}\!\otimes\!\hat{K}_{a_j,j}\,\hat{U}\ket{\Psi^-}
= \tfrac{1}{4\sqrt{2}}\!\left(a_i\, e^{i\xi_i} - a_j\, e^{i(\Theta_{ij}^{(\Psi)} + \xi_j)}\right),
\end{equation}
whose modulus squared gives the coincidence probability
\begin{equation}
P_{ij}^{(a_i a_j)} = \tfrac{1}{16}\!\left[1 - a_i a_j\, V \cos\!\left(\Theta_{ij}^{(\Psi)} - (\xi_i - \xi_j)\right)\right],
\label{eq:fringe-full}
\end{equation}
where the four port combinations sum to the central-slot coincidence probability $\tfrac{1}{4}$, and $V \leq 1$ is the heralded interferometric visibility. We assume $V=1$ in the ideal case derived here. $V$ is reduced in practice by source impurity, UMZI mismatch and residual polarization distinguishability. Opposite-port coincidences ($a_i a_j = -1$) track $1 + V\cos$, while same-port coincidences ($a_i a_j = +1$) track $1 - V\cos$; their normalized difference,
\begin{align}
E(\xi_i, \xi_j) &= \frac{P^{++} + P^{--} - P^{+-} - P^{-+}}{P^{++} + P^{--} + P^{+-} + P^{-+}} \nonumber \\ 
&= -V\cos\!\left(\Theta_{ij}^{(\Psi)} - (\xi_i - \xi_j)\right),
\label{eq:correlator}
\end{align}
is the two-photon correlator whose phase is the RF observable $\Theta_{ij}^{(\Psi)}$.

\subsection{Proposed network architecture}\label{sec:network}
Our proposed time-bin DQS network architecture is shown in Fig.~\ref{fig:network}. A time-bin Bell state is generated at the source node by first generating a polarization Bell state. This state is subsequently passed through a polarization-to-time-bin conversion module \cite{nehra2026}. This scheme allows us to generate both $\ket{\Psi^-}$ and $\ket{\Phi^+}$ states via standard waveplate rotations before mapping to the time-bin basis. The conversion module can include either a tunable or a fixed delay line, as discussed more below. Alternatively, one could implement a double-pump scheme with a pulse-shaper to generate time-bin entangled photon pairs directly, eliminating the need for a complex and lossy polarization-to-time converter at the source. The resulting limitation of only producing $\ket{EE} + \ket{LL}$ states is then bypassed at the sensor node by operating one of its RF phase modulators with flipped polarity ($\pi$ phase shift), effectively executing the required mode rotation directly during the transduction step. Either way, the signal and idler photons output from the source node are distributed to four remote sensor nodes through a set of beamsplitters, similar to the beamsplitter distribution network utilized in \cite{Kim2024}.

To achieve quadrature locking, one may implement the source node setup shown in Fig.~\ref{fig:network}(a), wherein a pulsed pump laser is triggered by an RF phase-locked loop. Alternatively, one could implement a pump-side stabilization scheme. For example, one could place a UMZI and a pulsed pump laser before the entanglement source, instead of after it. The UMZI will then generate the required Early-Late time-bin pump pulses. By utilizing both output ports of the UMZI, where one port is directed to the entanglement source and the other is routed to a feedback-loop photodetector locked to a centralized master clock, we can continuously track and compensate for phase drifts in the optical path \cite{Xavier2011}. This dual-port configuration provides a clean mechanism to stabilize the arrival time $t_e$ relative to the distributed RF reference without introducing loss or disruption to the primary sensing channel.

At the remote sensor nodes, depicted in Fig.~\ref{fig:network}(b), the incoming RF signal must be transduced onto the optical probe. This is achieved via an EOM or a similar microwave-to-optical transducer. To maximize the frequency agility of the network, we propose an architecture that mixes the incoming RF signal at carrier frequency $\omega_{\mathrm{RF}}$ down to an intermediate frequency (IF) using a local oscillator (LO). In this scheme, an antenna at each sensor node captures the RF field, which is then passed to an RF mixer alongside a tunable LO tone ($\omega_{\mathrm{LO}}$). The resulting IF signal, $\omega_{\mathrm{IF}} = |\omega_{\mathrm{RF}} - \omega_{\mathrm{LO}}|$, is fed into the EOM to modulate the arriving time-bin qubit. A similar antenna setup at the source node provides an absolute RF phase reference for the network.

Operating at an IF allows the network to maintain a fixed $\omega_{\mathrm{IF}}$ across a wide variety of target RF signal carrier frequencies by tuning $\omega_{\mathrm{LO}}$. Consequently, the UMZIs at the source and sensor nodes can be built with fixed, highly stable optical delay lines. In this LO-mixing configuration, the antipodal sampling condition becomes
\begin{equation} \label{eq:if}
    \omega_{\mathrm{IF}}\Delta \tau = \pi.
\end{equation} 
Here, $\Delta \tau$ is a static optical property of the network, and the classical $\omega_{\mathrm{LO}}$ is tuned to achieve the antipodal sampling condition for any given target $\omega_{\mathrm{RF}}$. For this method to work, the relative phase between distributed nodes must be preserved, i.e., the local oscillators across the network must be rigidly phase-locked.  If the independent LOs experience relative phase drift, that drift will be imprinted onto the EOM modulation, reducing or destroying the spatial correlation of the measured RF signal phase. The White Rabbit (WR) clock distribution can satisfy this requirement through low-jitter distribution of $10\;\mathrm{MHz}$ and $1$ pulse-per-second ($\mathrm{PPS}$) signals \cite{lipinski2011}. 

During the heterodyne down-conversion process, the low-pass filtered IF phase at node $j$ becomes 
\begin{equation}
\psi_{\mathrm{IF},j} = \psi_j - \psi_{\mathrm{LO},j} + \phi_{\mathrm{mix},j},
\end{equation}
where $\phi_{\mathrm{mix},j}$ represents RF mixing chain phase shifts. Because local oscillators across the network are phase-locked to the master clock reference ($\psi_{\mathrm{LO},i} = \psi_{\mathrm{LO},j}$), the relative LO phase vanishes in pairwise measurements. Any residual static offsets $(\phi_{\mathrm{mix},i} - \phi_{\mathrm{mix},j})$ can be removed via baseline calibration. Locking the LO at each sensor node to a centralized WR master clock also provides the absolute RF phase reference necessary to extract the sum-phase observable ($\ket{\Phi^+}$) and enables the quadrature-locked operation discussed below in Section~\ref{sec:results}. 

Alternatively, the network can be implemented as a direct-RF architecture, bypassing IF mixing entirely. In this configuration, the RF antenna signal drives the EOM directly, and $\Delta \tau$ is adjusted dynamically using tunable optical delay lines (e.g., switched fiber banks or free-space translation stages) within source and sensor node UMZIs. Direct RF operation may be desired if one wants to avoid $1/f$ phase noise near DC, or if one is operating in a congested RF spectrum where baseband mixing may introduce spurious signal overlap. Ultimately, the choice of architecture represents a design trade-off: operating at an IF places the phase-stabilization and tuning burden on classical RF hardware and accurate clock distribution, whereas direct-RF operation shifts that burden into the optical domain.

\begin{figure*}
  \centering
  \includegraphics[width=\linewidth]{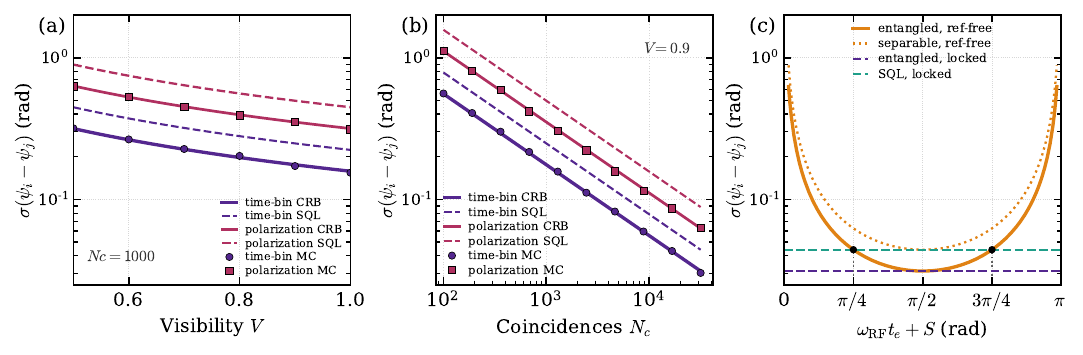} 
  \caption{ Metrology of the phase-difference observable using $\ket{\Psi^-}$ at $\varphi_0 = 0.1$. 
(a) Standard deviation of the phase-difference estimator versus interference visibility $V$ for $N_c = 1000$. 
(b) Standard deviation versus the number of coincidence events $N_c$ at $V = 0.9$ for time-bin (purple) and polarization (red) encodings. The time-bin curve sits $6\;\mathrm{dB}$ below the polarization curve due to the antipodal sub-cycle sampling advantage. Markers denote empirical Monte Carlo results over $R = 1,000$ trials. Standard errors on the sample standard deviation are smaller than the marker size. CRB: Cram\'er--Rao Bound; MC: Monte Carlo; SQL: standard quantum limit. 
(c) Standard deviation versus the common-mode phase envelope $\omega_{\mathrm{RF}}t_e + S$, where $S = (\psi_i+\psi_j)/2$. The reference-free standard deviation diverges at the envelope zeros, matches the phase-locked floor (purple dashed line) at quadrature phase $\pi/2$, and drops below the standard quantum limit (SQL) inside the $[\pi/4, 3\pi/4]$ operating window. The separable reference-free curve sits $3\;\mathrm{dB}$ above the entangled curve, demonstrating that the entanglement advantage persists even without phase locking. }
  \label{fig:kim_results}
\end{figure*}

\section{\label{sec:results}Results}

After the EOM action, the $\ket{\Psi^-}$ pair occupies the two-dimensional subspace spanned by $\{\ket{EL}, \ket{LE}\}$ and is, up to global phase, the pure state
\begin{equation}
\ket{\psi(\Theta)}_{ij} = \tfrac{1}{\sqrt{2}}\left(\ket{EL} - e^{i\Theta_{ij}^{(\Psi)}}\ket{LE}\right),
\end{equation}
a qubit carrying the relative phase $\Theta_{ij}^{(\Psi)}$. The QFI for a parameter $\theta$ encoded in a pure state is $F_Q = 4(\langle\partial_\theta\psi|
\partial_\theta\psi\rangle - |\langle\psi|\partial_\theta\psi\rangle|^2)$. Taking $\theta = \Theta_{ij}^{(\Psi)}$,
\begin{equation}
F_Q(\Theta_{ij}^{(\Psi)}) = 4\!\left(\tfrac{1}{2} - \big|\tfrac{-i}{2}\big|^2\right) = 4\!\left(\tfrac{1}{2} - \tfrac{1}{4}\right) = 1,
\end{equation}
the unit QFI of an ideal equatorial qubit phase. For a heralded detection fringe of visibility $V \leq 1$, the qubit is the mixed state 
\begin{align}
\rho(\Theta_{ij}^{(\Psi)}) = \frac{1}{2}(\mathbbm{1} - V \cos\Theta_{ij}^{(\Psi)}\,\hat{\sigma}_x - V \sin\Theta_{ij}^{(\Psi)}\,\hat{\sigma}_y),
\end{align}
whose Bloch vector $\vec{r} = (V\cos(\Theta_{ij}^{(\Psi)}), V\sin(\Theta_{ij}^{(\Psi)}), 0)$ has length $V$. The mixed state QFI formula yields the per-pair QFI 
\begin{equation}
    F_Q = |\partial_\Theta\vec{r}|^2 + (\vec{r}\cdot\partial_\Theta\vec{r})^2/(1-|\vec{r}|^2) = V^2
\end{equation}
which is consistent with the result for polarization Bell states in Kim et al. \cite{Kim2024}. While these QFI values match, the advantage of our dynamic architecture emerges when mapping the QFI back to the physical parameter of interest. 

In polarization-based DV-DQS networks, the estimand is the optical phase itself. Under our framework, the physical parameter is the RF carrier phase difference; for the $\Psi^-$ probe, $D_{ij} = \psi_i - \psi_j$. Estimating $D_{ij}$ requires traversing the EOM transduction mechanism. As we saw above, when operating at the quadrature bias point $\omega_{\mathrm{RF}}t_e = \pi/2$, the relative phase from Eq.~\ref{eq:rel_phase_psi} simplifies to $\Theta_{ij}^{(\Psi)} \approx 2\varphi_0(\psi_i - \psi_j) = 2\varphi_0 D_{ij}$ in the small-signal regime. See Appendix \ref{app:approx} for more details on this small-phase approximation. The sensitivity coefficient is therefore $\partial \Theta_{ij}^{(\Psi)} / \partial D_{ij} = 2\varphi_0$. This yields the per-pair QFI for the RF observable specifically: 
\begin{equation} 
F_Q^{\mathrm{ent}}(D_{ij}) = (2\varphi_0)^2\,V^2 = 4\,V^2\,\varphi_0^2. 
\label{eq:fqent} 
\end{equation}

Because the early and late time bins antipodally sample the RF wave, each photon acquires twice the phase of a standard 1:1 phase mapping. In the QFI, this coherent doubling yields a factor of 4—a \textbf{6 dB} structural enhancement over the factor of 1 achieved using polarization Bell states. 

The absolute precision is governed by the transducer prefactor $\varphi_0^2$. This factor, limited by the EOM half-wave voltage and antenna gain, represents the cost of mapping an external, dynamic RF field into the quantum optical domain. This does not negate the advantage of time-bin encoding, but makes measurement harder by limiting signal strength. 

We will now compare the above result to the analogous separable scenario. We define two separable single-photon probes (one per node), each prepared in the equal time-bin superposition state
\begin{equation}
    \ket{\Psi_{\mathrm{sep},k}} = \frac{1}{\sqrt{2}}\big( \ket{E_k} +  e^{i\phi_{0,k}}\ket{L_k} \big)
\end{equation}
with the same Early-Late bin separation $\Delta \tau$  that achieves the antipodal sampling condition. As each probe passes through its respective sensor node $k \in \{i,j\}$, it acquires a phase $\Delta\varphi_k = 2\varphi_0\psi_k$ via EOM transduction with visibility $V$. The total relative phase of the qubit becomes $\Phi_k = \phi_{0,k} + \Delta\varphi_k$. The associated single-node QFI evaluated with respect to the local RF phase $\psi_k$ is
\begin{equation}
    F_Q(\psi_k) = (d\Phi_k/d\psi_k)^2 V^2 = (d\Delta\varphi_k/d\psi_k)^2 V^2 = 4V^2\varphi_0^2,
\end{equation}
with variance $\mathrm{Var}(\psi_k) \ge 1/(4V^2\varphi_0^2)$. Taking the average between the two independent estimates at nodes $i,j$ gives
\begin{equation}
\mathrm{Var}(D_{ij}) = \mathrm{Var}(\psi_i) + \mathrm{Var}(\psi_j) = \frac{2}{4V^2\varphi_0^2} = \frac{1}{2V^2\varphi_0^2},
\end{equation}
which yields the separable QFI, for the exact same estimand, as
\begin{equation}
F_Q^{\mathrm{sep}}(D_{ij}) = 2\,V^2\,\varphi_0^2.
\label{eq:fqsep}
\end{equation}
The Cram\'er--Rao variance ratio is therefore
\begin{equation}
\frac{\sigma^2_{\mathrm{ent}}}{\sigma^2_{\mathrm{sep}}} = \frac{F_Q^{\mathrm{sep}}}{F_Q^{\mathrm{ent}}} = \frac{2V^2\varphi_0^2}{4V^2\varphi_0^2} = \frac{1}{2},
\label{eq:dbratio}
\end{equation}
confirming the $3$~dB quantum advantage per coincidence event over the SQL. Eq.~\ref{eq:dbratio}  shows that the entanglement advantage  over separable time-bin probes is independent of the visibility $V$ and the modulation depth $\varphi_0^2$.  This is the $N= 2$ Heisenberg limit for the two-photon probe and matches the per-pair scaling of Liu et al. \cite{Liu2021}.

\begin{figure}[htbp]
  \centering
  \includegraphics[width=\linewidth]{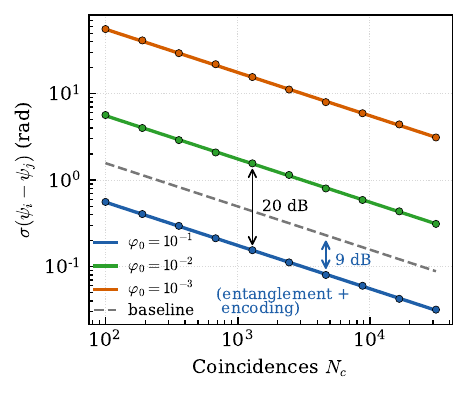} 
  \caption{The cost of RF-to-optical transduction. The gray baseline curve is the SQL for separable operation with polarization Bell states and $\varphi_0=0.1$. The blue curve also corresponds to a modulation depth of $\varphi_0=0.1$ for entangled time-bin operation, which was used in Fig.~\ref{fig:kim_results}. If the modulation depth is reduced by a factor of $10$, the standard deviation jumps by $20\;\mathrm{dB}$ (green curve, another jump for the orange curve). The reduction in precision can be offset somewhat by order-of-magnitude increases in the number $N_c$ of two-photon coincidence events.
 }
  \label{fig:transduction}
\end{figure}

We simulate the performance of our method using the phase-difference observable and the $\ket{\Psi^-}$ Bell state as a probe, with results shown in Fig.~\ref{fig:kim_results}. Assuming a modulation depth of $\varphi_0 = 0.1$ and $N_c=1000$ coincident detection events, we plot the standard deviation of the difference observable estimate as a function of the interferometric visibility $V$ in Fig.~\ref{fig:kim_results}(a). The standard deviation is also plotted as a function of $N_c$ in Fig.~\ref{fig:kim_results}(b) for $V=0.9$. A maximum likelihood estimator applied to the simulated coincidence data saturates the time-bin Cram\'er--Rao bound (CRB) for visibilities down to $V=0.5$ and across three orders of magnitude in the number of coincidences $N_c$, confirming that the per-pair QFI is achievable under finite statistics. The entangled CRB sits $3\;\mathrm{dB}$ below the separable SQL for both the polarization and time-bin encodings. The time-bin and polarization curves are separated by $6\;\mathrm{dB}$, reflecting the standard deviation reduction enabled by antipodal sub-cycle sampling within the time-bin encoding. The four curves are computed independently from their respective Fisher information formulas, and markers show the corresponding Monte Carlo estimates. See Appendix \ref{app:montecarlo} for simulation details. 

Under non-locked operation, $t_e$ is uncontrolled relative to the RF carrier, so the common-mode envelope $\sin(\omega_{\mathrm{RF}}t_e+S)$ takes values across its full range from shot to shot. Quadrature locking so that $\omega_{\mathrm{RF}}t_e=\pi/2$ allows one to place every photon at the envelope optimum, recovering the full per-coincidence sensitivity for the phase-difference observable. This requires synchronizing photon emission to the RF carrier phase. To achieve this, one can lock the probe emission time to a reference antenna that also detects the RF carrier so that photons are always emitted at a controlled RF phase value. Fixed geometric phase offsets picked up by the probe photons as they traverse the network can be folded into the same common-mode parameter $S=\frac{\psi_i + \psi_j}{2}$.

Absent the quadrature lock, $t_e$ can instead be timestamped against a distributed network clock (e.g. White Rabbit), and the envelope removed in post-processing. This demodulate-in-post mode of operation eliminates the need for RF-carrier-phase synchronization at the cost of a factor-$\sqrt{2}$ reduction in precision arising from the envelope average $\langle \sin^2(\omega_{\mathrm{RF}}t_e+S) \rangle=1/2$. 
Alternatively, one may post-select coincidences whose tagged arrival phase falls near the envelope optimum, recovering close to the full per-coincidence sensitivity on the retained shots. This option trades the $\sqrt{2}$ precision penalty for a reduction in the usable coincidence rate, with the two strategies yielding equivalent precision per unit integration time.

The impact of quadrature-locked versus reference-free operation is explored in Fig.~\ref{fig:kim_results}(c), where reference-free is understood to mean unlocked, timestamped operation. The standard deviation of the reference-free curve follows the $\sin \big( \omega_{\mathrm{RF}}t_e  + S \big)$ modulation, as expected, and the reference-free bound for the separable case sits $3\;\mathrm{dB}$ higher. Reference-free entangled operation can achieve sub-SQL performance when the envelope phase sits inside the $[\pi/4, 3\pi/4]$ range, which spans half of the possible arrival phases, so that even without a phase reference, the protocol outperforms reference-locked separable probes over a substantial fraction of possible operating conditions. Outside of this window, the reference-free curve standard deviation sits above the locked SQL, and the envelope-averaged precision sits exactly on it.

Finally, to quantify the impact of low-efficiency microwave-optical transduction, we simulate the performance of the network at a range of transduction efficiencies. Fig.~\ref{fig:transduction} plots the achievable measurement precision versus the number of coincident detection events $N_c$ for modulation depths $\varphi_0 = 10^{-1}, 10^{-2}, 10^{-3}$. The dashed line in the plot corresponds to the baseline performance of a measurement utilizing polarization Bell states. A $10\times$ reduction in modulation depth comes with a $20\;\mathrm{dB}$ precision penalty at a given $N_c$. While severe, the loss of precision can be mitigated. For example, it can be overcome by increasing $N_c$ by $100\times$. As entanglement sources mature and continue to become brighter \cite{altepeter2005, steiner2021, yadav2024, zhang2025}, this mitigation strategy is realistic. Furthermore, ongoing advances in integrated EOMs \cite{wang2018, kharel2021, yue2024} could significantly improve the single-pass modulation depth in future experimental demonstrations.

\section{Discussion} \label{sec:discussion}

We have proposed and analyzed a DV-DQS method for RF sensing based on time-bin encoded probe states. Here, the demodulation reference itself is an intrinsic aspect of the probe state, as opposed to part of a classical back-end. This is possible because the time-bin separation $\Delta \tau$ of a time-bin Bell state provides the internal temporal structure that polarization encoding lacks. The demodulation identity (Eq.~\ref{eq:demod_identity}) converts the probe into a coherent sub-cycle sampler of the RF carrier phase. While previous work has leveraged the temporal structure of a quantum probe state for time of arrival localization \cite{he2024}, ours is the first proposal to apply this structure directly to RF sensing. Time-bin DV-DQS can be applied to e.g. Angle of Arrival (AoA) geolocation measurements - see Appendix \ref{app:AoA} for more details. 

Beyond the $6\;\mathrm{dB}$ precision gain enabled by coherent doubling, time-bin qubits carry the additional advantage of protection from environmental perturbations. Their relative insensitivity to polarization mode dispersion \cite{marcikic2004, knaut2024} provides resilience against phase drifts that would otherwise disturb the RF phase measurement, which holds promise for real-world operation. Inefficient microwave-optical transduction is a significant challenge limiting the achievable modulation depth, but can be at least partially overcome by using brighter entanglement sources. Finite time-bin widths and separation mismatches have a mild impact on the achievable modulation depth at the sensors, as detailed in Appendices \ref{app:bin width}, \ref{app:bin mismatch}.

The proposed network can be operated at the $\Delta \tau \omega_{\mathrm{RF}} = \pi$ setpoint or at an IF via Eq.~\ref{eq:if}.  Both options allow for high-speed $\Delta \tau$ tuning for agile operation across the spectrum, either through a tunable UMZI delay line or a tunable RF local oscillator. The network can be operated across a large range of RF carrier frequencies as well. The maximum frequency that can be demodulated with the qubit will be determined by the timing resolution of the single photon detectors, since this jitter sets a lower bound on the resolvable $\Delta \tau$. If we assume that the SNSPDs used in the network have a $20\;\mathrm{ps}$ jitter, we will require a minimum $40-50 \;\mathrm{ps}$ time-bin separation, resulting in a theoretical $10-12\;\mathrm{GHz}$ (X-band) maximum detectable carrier frequency. The maximum IF frequency may be capped further by timing jitter associated with phase-locking and distributing the LO. If we assume a White Rabbit timing jitter of $20-50\;\mathrm{ps}$, the associated phase jitter induces a mild visibility penalty ($\approx 1\%$) at an IF of $500\;\mathrm{MHz}$, with phase jitter becoming a dominant degradation factor as the IF approaches the low-GHz regime.

On the other hand, the minimum frequency achievable without IF mixing will be set by the maximum time delay in a UMZI that can be phase-stabilized. If we estimate that the maximum time delay that can be stabilized is $\Delta \tau = 1 \;\mathrm{\mu s}$, that corresponds to a minimum frequency of $500\;\mathrm{kHz}$ (Medium Frequency band). 

While we model the incoming RF signal as a continuous wave (CW) here, the proposed network can theoretically detect any RF waveform, e.g. CW, pulsed, or phase-keyed, provided the carrier phase is stable during $\Delta \tau$. The minimum detectable RF temporal pulse duration is limited in practice, however. Extracting an RF phase difference with precision set by $\sigma_D$ (see Appendix \ref{app:AoA}) requires accumulating 
\begin{equation}
    N_c = \frac{1}{4 \sigma_D^2 \varphi^2_0V^2}
\end{equation}
coincident detection events to build the necessary interference fringe and apply a maximum likelihood estimator, since a single photon pair provides only a single detection event at most. If the RF signal of interest is a single, non-repeating pulse, this pulse must remain on long enough for the network to generate, distribute, and detect $N_c$ coincidences. The shortest measurable pulse duration is therefore fundamentally limited by the brightness of the entanglement source and the end-to-end transmission efficiency of the network. 

This minimum required pulse duration also scales inversely with the signal strength: since a 10-fold reduction in modulation depth $\varphi_0$ requires a 100-fold increase in the number of coincidences to maintain the same precision, a weaker pulse must necessarily last significantly longer than a strong pulse to be successfully detected. For example, if we aim to detect a signal within a phase deviation of $\sigma_D = 0.1\;\mathrm{rad}$ and we have $\varphi_0 = 0.1$ and $V=0.9$, then we require $N_c=3,086 $ coincident detection events to achieve the target precision. If we are detecting a phase keyed signal with a $1\;\mathrm{kHz}$ modulation frequency, then we must accumulate $N_c$ coincident detection events within a $1\;\mathrm{ms}$ integration time. This requirement, though stringent, is possible to satisfy with contemporary bright entanglement sources that mitigate two-photon absorption effects \cite{steiner2021} and commercial single photon detectors, as long as the network loss is low and the modulation depth is high.

We have proposed and theoretically characterized a new class of DV-DQS protocols which leverage time-bin encodings to sense RF fields. Time-bin encoding grants the DQS network access to non-static fields and offers a significant sensitivity improvement when properly calibrated and locked. The factor of $2$ increase in precision represents a $6\;\mathrm{dB}$ gain from antipodal two-time sampling relative to a single-pass static mapping. Entanglement independently provides a $3\;\mathrm{dB}$ precision advantage over separable time-bin probes. We show that there is an asymmetry in pairwise phase sum and difference observables, with phase differences being easier to extract  from measurements. The proposed method leverages mature, existing RF and quantum photonic technologies, and near-term experimental implementations are feasible. It also opens up a new field for DQS research with time-bin encoded probes. Future theoretical and experimental work can explore a variety of discrete and continuous variable time-domain probes, such as time-bin $N00N$ or cluster states. 


\section{Author contributions}
All authors contributed to the development of the theoretical protocol and proposed network architecture. E.S. performed theoretical simulations. All authors contributed to manuscript drafting. E.S., B.K. and D.H. directed and supervised the project.

\section{Conflicts of interest}
The authors declare no conflicts of interest.

\section{Acknowledgments}
The authors gratefully acknowledge funding from the Microelectronics Commons program through the NORDTECH hub. This work was funded in part by the Air Force Research Laboratory (AFRL) under contract \#FA8750-26-C-B045. Any opinions, findings, and conclusions or recommendations expressed in this article are those of the authors and do not necessarily reflect the views of the Air Force Research Laboratory (AFRL). 

\def\bibsection{\section*{\centering References}}
\bibliography{library}
\newpage

\appendix

\section{Quantum Fisher Information Beyond the Small-Phase Approximation}\label{app:approx}

\noindent Beyond the small-phase approximation, the Bell-state phase depends nonlinearly upon both collective RF phase coordinates,

\begin{equation}
\Theta(\nu_+,\nu_-)
=
4\varphi_0
\sin\!\left(
\omega_{\mathrm {RF}}t_e+\frac{\nu_+}{2}
\right)
\sin\!\left(
\frac{\nu_-}{2}
\right),
\end{equation}

\noindent where $\nu_\pm=\psi_1\pm\psi_2$. Since this RF phase dependency occurs only through the single parameter $\Theta$, the chain rule gives

\begin{equation}
(F_\nu)_{ab}
=
F_Q(\Theta)
\frac{\partial\Theta}{\partial\nu_a}
\frac{\partial\Theta}{\partial\nu_b}=V^2
\frac{\partial\Theta}{\partial\nu_a}
\frac{\partial\Theta}{\partial\nu_b},
\end{equation}

\noindent using $F_Q(\Theta)=V^2$, derived in the main text of the paper.  At the quadrature operating point, $\omega_{\mathrm{RF}}t_e=\pi/2$, the exact collective RF phase QFIM becomes

\begin{equation}
\begin{split}
    F_\nu = 4V^2\varphi_0^2
    \begin{pmatrix}
        \sin^2\!\left(\dfrac{\nu_+}{2}\right) \sin^2\!\left(\dfrac{\nu_-}{2}\right) & 
        \begin{split}
            -\sin\!\left(\dfrac{\nu_+}{2}\right) \cos\!\left(\dfrac{\nu_+}{2}\right) \\
            \times\sin\!\left(\dfrac{\nu_-}{2}\right) \cos\!\left(\dfrac{\nu_-}{2}\right)
        \end{split} \\
        \\
        \begin{split}
            -\sin\!\left(\dfrac{\nu_+}{2}\right) \cos\!\left(\dfrac{\nu_+}{2}\right) \\
            \times\sin\!\left(\dfrac{\nu_-}{2}\right) \cos\!\left(\dfrac{\nu_-}{2}\right)
        \end{split} & 
        \cos^2\!\left(\dfrac{\nu_+}{2}\right) \cos^2\!\left(\dfrac{\nu_-}{2}\right)
    \end{pmatrix}
\end{split}
\end{equation}

\noindent Since this matrix is an outer product, it possesses a single non-zero eigenvalue,

\begin{equation}
\lambda_1
=
4V^2\varphi_0^2
\left[
\sin^2\!\left(\frac{\nu_+}{2}\right)
\sin^2\!\left(\frac{\nu_-}{2}\right)
+
\cos^2\!\left(\frac{\nu_+}{2}\right)
\cos^2\!\left(\frac{\nu_-}{2}\right)
\right],
\end{equation}

\noindent with corresponding eigenvector

\begin{equation}
\mathbf c_1
\propto
\begin{pmatrix}
-\sin\!\left(\dfrac{\nu_+}{2}\right)
\sin\!\left(\dfrac{\nu_-}{2}\right)
\\[2mm]
\cos\!\left(\dfrac{\nu_+}{2}\right)
\cos\!\left(\dfrac{\nu_-}{2}\right)
\end{pmatrix},
\end{equation}

\noindent which defines the optimal sensing direction.  Thus, beyond the small-phase approximation, the optimally sensed parameter is generally a modulation-dependent combination of the RF phase sum and difference. Expanding about the origin gives

\begin{equation}
\hat{\mathbf c}_1
=
\hat{\boldsymbol{\nu}}_-
-
\frac{\nu_+\nu_-}{4}
\hat{\boldsymbol{\nu}}_+
+
O(\nu^4),
\end{equation}

\noindent showing that the optimal sensing direction remains aligned with $\nu_-$ to first order, with the leading rotation appearing only through the mixed second-order term $\nu_+\nu_-$.  The Quantum Fisher Information associated specifically with estimating the differential RF phase is

\begin{equation}
F_{\nu_-\nu_-}
=
4V^2\varphi_0^2
\cos^2\!\left(\frac{\nu_+}{2}\right)
\cos^2\!\left(\frac{\nu_-}{2}\right),
\end{equation}

\noindent whose second-order expansion is

\begin{equation}
F_{\nu_-\nu_-}
=
4V^2\varphi_0^2
\left[
1-
\frac{\nu_+^2+\nu_-^2}{4}
\right]
+
O(\nu^4),
\end{equation}

\noindent yielding the differential RF phase uncertainty

\begin{equation}
\Delta\nu_-
\ge
\frac{1}{2\varphi_0V\sqrt N_c}
\left[
1+
\frac{\psi_1^2+\psi_2^2}{4}
+
O(\psi^4)
\right].
\end{equation}

\noindent  Thus, finite modulation gradually rotates the optimal sensing direction away from the RF phase difference while simultaneously reducing the differential-phase Quantum Fisher Information. Both effects, however, first appear at second order, demonstrating the perturbative robustness of the small-phase approximation around the origin.

\section{Impact of finite bin width}\label{app:bin width}
The phase measurement result in Main Text Eq.~\eqref{eq:early_late_gain} treats each time-bin as
instantaneous. In reality, a physical time-bin is a wave packet of finite duration. The wave packet can be  described by a normalized intensity envelope $|f(u)|^2$ centered on its arrival time ($t_e$ for the Early bin and $t_l$ for the Late bin), with $\int |f(u)|^2\,du = 1$. During RF phase sensing, an EOM imprints the instantaneous RF phase $\varphi_j(t)$ across the non-instantaneous wave packet. Assuming that we are working again in the small-signal regime $\varphi_0 \ll 1$, the Early time-bin acquires the envelope-averaged phase
\begin{equation}
\langle\varphi_e^j\rangle = \int |f(u)|^2\,\varphi_0\cos\!\big(\omega_{\mathrm{RF}}(t_e + u) + \psi_j\big)\,du.
\end{equation}
Expanding the cosine term and using the evenness of a symmetric envelope (so the $\sin\omega_{\mathrm{RF}} u$ term integrates to zero) gives
\begin{equation}
\langle\varphi_e^j\rangle = \varphi_0\, \mathcal{F}(\omega_{\mathrm{RF}})\, \cos(\omega_{\mathrm{RF}} t_e + \psi_j),
\end{equation}
where $\mathcal{F}$ is the Fourier cosine transform of the intensity envelope,
\begin{equation}
\mathcal{F}(\omega_{\mathrm{RF}}) \equiv \int |f(u)|^2 \cos(\omega_{\mathrm{RF}} u)\,du = \mathrm{Re}\!\int |f(u)|^2 e^{i\omega_{\mathrm{RF}} u}\,du.
\label{eq:formfactor}
\end{equation}
The Late time-bin, sampled a half-period later under $\omega_{\mathrm{RF}}\Delta \tau = \pi$, acquires the phase
\begin{equation}   
\langle\varphi_l^j\rangle = -\varphi_0\mathcal{F}(\omega_{\mathrm{RF}})\cos(\omega_{\mathrm{RF}} t_e + \psi_j),
\end{equation} so the overall relative phase imprinted on the qubit becomes

\begin{equation}
\Delta\varphi_j = -2\varphi_0\,\mathcal{F}(\omega_{\mathrm{RF}})\cos(\omega_{\mathrm{RF}} t_e + \psi_j).
\end{equation}
The finite time-bin width therefore acts as an effective reduction of the sensor's modulation depth, $\varphi_0 \to \varphi_0\,\mathcal{F}(\omega_{\mathrm{RF}})$. The reduced modulation depth enters the per-pair quantum Fisher information~\eqref{eq:fqent} as a factor of $\mathcal{F}^2$ multiplying $\varphi_0^2$. Wave packet distortion induced by the RF phase modulation, which would degrade the fringe visibility proper, is $\mathcal{O}(\varphi_0^2)$ and negligible in the linear regime.

For Gaussian time-bin envelope with RMS width $\sigma$,
\begin{equation}
    |f(u)|^2 = (2\pi\sigma^2)^{-1/2}\exp(-u^2/2\sigma^2),
\end{equation}
we have
\begin{equation}
\mathcal{F}_{\mathrm{gauss}}(\omega_{\mathrm{RF}}) = \exp\!\left(-\tfrac{1}{2}\omega_{\mathrm{RF}}^2\sigma^2\right).
\label{eq:gauss}
\end{equation}

For a Gaussian bin of rms width $\sigma = 20$~ps at $\omega_{\mathrm{RF}}/(2\pi) = 3$~GHz, the argument is $\omega_{\mathrm{RF}}\sigma = 0.377$~rad and $\mathcal{F}_{\mathrm{gauss}} = 0.931$, which causes a $6.9\%$ reduction in effective modulation depth and a $13.3\%$ reduction in per-pair quantum Fisher information.

The Fourier term $\mathcal{F}$ sets a soft upper bound on the usable bin width. Achieving $\mathcal{F} \geq 0.9$ requires $\omega_{\mathrm{RF}}\sigma \lesssim 0.459$, or $\sigma \lesssim 0.073/f_{\mathrm{RF}}$ for the Gaussian case. At $3$~GHz this constraint permits bin widths of up to $\sigma \sim 24$~ps. Operation at higher $\omega_{\mathrm{RF}}$ tightens this bound proportionally, whereas operation at an intermediate frequency (IF) $\omega_{\mathrm{IF}}= 300 \;\mathrm{MHz}$ within a network using local oscillators at each sensor permits bin widths of up to $\sigma \sim 243$~ps. The additional flexibility in bin widths provides an additional motivation for utilizing local oscillators within the network and mixing to baseband, as discussed in the Main Text. 

Fiber chromatic dispersion causes wavepacket broadening $\sigma(L)$ of the form
\begin{equation}
    \sigma(L)=\sqrt{\sigma_0^2+(D\cdot \Delta \lambda \cdot L)^2}
\end{equation}
with an initial root-mean square time-bin width $\sigma_0^2$, a chromatic dispersion parameter $D$ in $\mathrm{ps/(nm\cdot km)}$, a single-photon wavepacket spectral bandwidth $\Delta \lambda$, and a fiber transmission length $L$. This effect imposes a maximum fiber length that can be accommodated before dispersion compensation is required to maintain $\mathcal{F} \geq 0.9$.

\section{Impact of time-bin mismatch}\label{app:bin mismatch}
When $\Delta \tau$ is mismatched to the true emitter frequency by $\delta$ such that $\omega_{\mathrm{RF}} \Delta \tau = \pi + \delta$, the Early and Late time bins no longer sample the carrier at exactly antipodal phases. The qubit relative phase is
\begin{equation} \delta\varphi_j(\delta) = \varphi_0\big[\cos(\omega_{\mathrm{RF}}(t_e+\Delta \tau)+\psi_j) - \cos(\omega_{\mathrm{RF}} t_e + \psi_j)\big], \label{eq:offres-raw} \end{equation} 
and substituting $\omega_{\mathrm{RF}}\Delta \tau = \pi + \delta$ then simplifying yields 
\begin{equation} \delta\varphi_j(\delta) = -2\varphi_0 \cos(\delta/2)\,\cos(\omega_{\mathrm{RF}} t_e + \psi_j + \delta/2). \label{eq:offres} 
\end{equation}
The on-resonance result (Eq.~\ref{eq:early_late_gain}) is recovered at $\delta = 0$. Coincidence phase sensitivity rolls off with $\delta$ as $\cos(\delta/2)$, and the apparent carrier phase rotates by $\delta/2$. The roll-off is quadratic near the operating point [$\cos(\delta/2)\approx 1-\delta^2/8$], so fringe visibility is first-order insensitive to small  $\Delta \tau$ detunings, with full nulls only at $\delta = \pi$ (i.e. at $\omega_{\mathrm{RF}}$ mismatched by a factor of $2$). This flexibility is operationally useful when emitter carrier frequencies are bracketed but not precisely known. The same roll-off governs fixed-IF operation, but with a wider tolerance. Suppose the carrier is down-converted to a fixed intermediate frequency $\omega_{\mathrm{IF}}$ and the sensor's UMZI satisfies $\omega_{\mathrm{IF}}\Delta \tau = \pi$.  A shift $\Delta f$ in the carrier frequency passes through the mixer to the IF, producing an IF-referenced detuning of 

\begin{equation}
\delta = 2\pi\,\Delta f\,\Delta \tau = \pi\,(\Delta f/f_{\mathrm{IF}})
\end{equation}
where $f_{\mathrm{IF}} = \omega_{\mathrm{IF}}/2\pi$. The half-contrast points $\delta = \pm\pi/2$ occur at values $\Delta f = \pm f_{\mathrm{IF}}/2$. This value is a fixed capture bandwidth of the sensor set by the IF alone, rather than a fixed fraction of the carrier.  Because this range is independent of where in the RF band the carrier sits, a single fixed interferometer retains the graceful roll-off of Eq.~\eqref{eq:offres} across the entire tunable range of operation.

\section{Monte Carlo simulations} \label{app:montecarlo}
To numerically validate the theoretical precision bounds, Monte Carlo simulations modeled the coincidence measurements as independent probabilistic trials. For a given number of registered coincident detection events $N_c$, the binary analyzer outcomes $y_n \in \{0, 1\}$ follow a Bernoulli distribution defined by the system's interference fringe:
\begin{equation}
    y_n \sim \mathrm{Bernoulli}(P_n), \quad \text{with} \quad P_n = \frac{1}{2}\left[1 + V \cos\left(\Theta_n - \xi\right)\right]
\end{equation}
where $V$ is the visibility, $\xi$ is the analyzer phase (set to $\pi/2$ for optimal quadrature operation), and $\Theta_n$ is the state-dependent imparted optical phase. 

For a static, locked parameter estimation, the aggregated coincidences $K = \sum_{n=1}^{N_c} y_n$ naturally follow a Binomial distribution, $K \sim \mathrm{Binomial}(N_c, P)$. The target phase difference $\Delta\psi = \psi_i - \psi_j$ is reconstructed via the non-linear inverse mapping:
\begin{equation}
\Delta\hat{\psi} = \frac{1}{k \varphi_0} \arcsin\!\left( \frac{2(K/N_c) - 1}{V} \right)
\end{equation}
where $k$ denotes the specific encoding sensitivity factor ($k=2$ for time-bin antipodal sampling, and $k=1$ for polarization). 

For estimation over dynamic or unlocked reference phases, the parameters are extracted using a linearized first-order maximum-likelihood estimator with local weights $w_n = \partial \Theta_n / \partial \Delta\psi$:
\begin{equation}
    \Delta\hat{\psi} = \frac{\sum_{n=1}^{N_c} w_n \left(y_n - \frac{1}{2}\right)}{\frac{V}{2} \sum_{n=1}^{N_c} w_n^2}
\end{equation}

The achievable physical precision is then quantified by evaluating the sample variance of these estimates over $R$ independent repetitions of the simulation:
\begin{equation}
    \mathrm{Var}(\Delta\hat{\psi}) = \frac{1}{R-1} \sum_{r=1}^R \left( \Delta\hat{\psi}_r - \langle \Delta\hat{\psi} \rangle \right)^2
\end{equation}
In our simulations, this empirical variance is directly compared against the analytical Cram\'er--Rao bounds to verify optimal saturation.

\section{Metrology of the phase sum observable}
While the main text focuses on the phase difference observable extracted using the $\ket{\Psi^-}$ probe, our architecture can also estimate the network-averaged antenna phase by probing with the $\ket{\Phi^+}$ state. We define this estimand as $\bar{\psi} = (\psi_i + \psi_j)/2$, following standard distributed quantum sensing conventions \cite{Liu2021}. 

After the EOM action, the $\ket{\Phi^+}$ pair occupies the two-dimensional subspace spanned by $\{\ket{EE}, \ket{LL}\}$ and is, up to global phase, the pure state
\begin{equation}
\ket{\psi(\Theta)}_{ij} = \tfrac{1}{\sqrt{2}}\left(\ket{EE} + e^{i\Theta_{ij}^{(\Phi)}}\ket{LL}\right),
\end{equation}
a qubit carrying the relative phase $\Theta_{ij}^{(\Phi)} = \Delta\varphi_i + \Delta\varphi_j$. Following the exact same geometric mixed-state QFI derivation utilized in the main text, a heralded detection fringe of visibility $V \leq 1$ yields a per-pair QFI of
\begin{equation}
F_Q(\Theta_{ij}^{(\Phi)}) = V^2.
\end{equation}
The distinction between the sum and difference observables arises when mapping this QFI back to the physical RF parameter via the chain rule. At quadrature bias ($\omega_{\mathrm{RF}}t_e = \pi/2$), the relative phase simplifies to
\begin{equation}\Theta_{ij}^{(\Phi)} \approx 2\varphi_0(\psi_i + \psi_j) = 4\varphi_0\bar{\psi}.\end{equation} 
The sensitivity coefficient is therefore $\partial\Theta_{ij}^{(\Phi)}/\partial\bar{\psi} = 4\varphi_0$  yielding
\begin{equation}
F_Q^{\mathrm{ent}}(\bar{\psi}) = (4\varphi_0)^2\,V^2 = 16\,V^2\,\varphi_0^2.
\label{eq:fqent_sum}
\end{equation}
As in the Main Text, we compare this to a separable strategy where two independent single-photon probes (one per node), each a local equatorial qubit of visibility $V$, acquire a phase $\Delta\varphi_k = 2\varphi_0\psi_k$. These probes estimate their local phases with a single-node QFI of $F_Q(\psi_k) = (\partial\Delta\varphi_k/\partial\psi_k)^2 V^2 = 4V^2\varphi_0^2$, hence a variance of $\mathrm{Var}(\psi_k) \geq 1/(4V^2\varphi_0^2)$. Forming the average from these independent estimates yields a variance of
\begin{equation}
\mathrm{Var}(\bar{\psi}) = \tfrac{1}{4}\!\left[\mathrm{Var}(\psi_i) + \mathrm{Var}(\psi_j)\right] = \frac{1}{8V^2\varphi_0^2},
\end{equation}
so that the separable QFI for the averaged phase observable is
\begin{equation}
F_Q^{\mathrm{sep}}(\bar{\psi}) = 8\,V^2\,\varphi_0^2.
\label{eq:fqsep_sum}
\end{equation}

The Cram\'er--Rao variance ratio is therefore
\begin{equation}
\frac{\sigma^2_{\mathrm{ent}}}{\sigma^2_{\mathrm{sep}}} = \frac{F_Q^{\mathrm{sep}}}{F_Q^{\mathrm{ent}}} = \frac{8V^2\varphi_0^2}{16V^2\varphi_0^2} = \frac{1}{2}.
\end{equation}
This confirms that the $3$~dB quantum advantage (the $N=2$ Heisenberg limit) is perfectly preserved for the sum-phase observable. 

As noted in Section~\ref{sec:results}, extracting this sum-phase parameter requires the global envelope $\sin(\omega_{\mathrm{RF}}t_e + S)$ to be strictly controlled. Therefore, while the $\ket{\Phi^+}$ protocol enjoys the same $3$~dB advantage and coherent doubling enhancement as the $\ket{\Psi^-}$ protocol, it imposes a stricter physical requirement on the network: the source node must share a rigid, absolute RF phase reference with the distributed RF fields to successfully lock the arrival time $t_e$.

\section{Application to Angle of Arrival estimation}\label{app:AoA}

The time-bin distributed quantum RF sensing network can be applied to angle-of-arrival (AoA) measurements, as has been explored elsewhere \cite{Xia2020, Sun2022, Li2023}. Here, we illustrate how our protocol maps onto this specific use case. 

If the sensor nodes are arranged in a uniform linear array with spacing $d$, an incoming plane wave with wavelength $\lambda$ at an incident angle $\theta_{\mathrm{AoA}}$ imparts a phase
\begin{equation}
    \psi_j = \frac{2 \pi d}{\lambda}j \sin(\theta_{\mathrm{AoA}}),
\end{equation}
which, for a half-wavelength spacing $d=\lambda/2$, becomes
\begin{equation}
    \psi_j = \pi j \sin(\theta_{\mathrm{AoA}}).
\end{equation}
If we probe the sensors with a $\ket{\Psi^-}$ time-bin Bell state, we will measure the pairwise phase difference between two sensors $i,j$:
\begin{equation}
    D_{ij} \equiv \psi_i - \psi_j = \pi (i-j)\sin(\theta_{\mathrm{AoA}}).
\end{equation}
The true AoA is determined through inversion of this observable:
\begin{equation}
    \hat{\theta}_{\mathrm{AoA}}=\arcsin \left( \frac{D_{ij}}{\pi (i-j)}\right).
\end{equation}
We calculate the angular precision by starting with the per-pair quantum Fisher information. Applying the chain rule to the relative phase $\Theta_{ij}^{(\Psi)} \approx 2\varphi_0 D_{ij}$, the pair estimates $D_{ij}$ with a QFI of
\begin{equation}
    F_Q(D_{ij}) = \left(\frac{\partial\Theta_{ij}^{(\Psi)}}{\partial D_{ij}}\right)^2 F_Q(\Theta) = 4\varphi_0^2 V^2
\end{equation}
per coincidence. After $N_c$ coincident detection events, the Cram\'er--Rao bound is
\begin{equation}
    \mathrm{Var}(D_{ij}) = \frac{1}{N_c F_Q(D_{ij})} = \frac{1}{4N_c\varphi_0^2 V^2}.
\end{equation}
Taking the square root, the phase-difference standard deviation is
\begin{equation}
    \sigma_D = \frac{1}{2\varphi_0 V \sqrt{N_c}}.
\end{equation}
Propagating the standard deviation through the angular inversion produces an angular precision of
\begin{align}
    \sigma_{\theta_{\mathrm{AoA}}} = \left|\frac{d\hat{\theta}_{\mathrm{AoA}}}{dD_{ij}}\right|\sigma_D = \frac{\sigma_D}{\pi(i-j) \cos(\theta_{\mathrm{AoA}})}\\ = \frac{1}{2\pi(i-j) \varphi_0 V \sqrt{N_c}\cos(\theta_{\mathrm{AoA}})}.
\end{align}
The $1/\cos(\theta_{\mathrm{AoA}})$ term corresponds to the standard geometric dilution of precision for a uniform linear sensor array, and the $\varphi_0V$ term carries the dependence on the transduction depth and visibility. The $3\;\mathrm{dB}$ time-bin encoding advantage is inherited directly through $F_Q$. Measurement precision improves as the baseline increases, so the sensor pair with the largest separation will return the most precise single-pair angle estimate. Multiple sensor pair baselines can be combined using a weighted least squares fit of the measured $D_{ij}$ to the unknown $\sin(\theta_{\mathrm{AoA}})$. While this mapping is illustrative of one specific sensor deployment, the protocol and its metrology are independent of any specific sensor geometry.

\end{document}